\documentclass[aps,pre,groupedaddress,amssymb,amsmath,pra,floatfix,showpacs]{revtex4}
\usepackage{epsfig}

\begin{document}
\title{Recovery of the starting times of delayed signals}
\author{L. Perotti, D. Vrinceanu, and D. Bessis}
\affiliation{Department of Physics, Texas Southern University, Houston, Texas\\
77004 USA}
\date{\today}

\begin{abstract}

We present a new method to locate the starting points in time of an arbitrary number of (damped) delayed signals. For a finite data sequence, the method permits to first locate the starting point of the component with the longest delay, and then --by iteration-- all the preceding ones. Numerical examples are given and noise sensitivity is tested for weak noise.

\end{abstract}

%\pacs{07.05.Kf, 07.05.Rm, 02.70.Hm}

\maketitle

\section{Introduction}

In signal processing we often encounter situations where the onset of the signal --or of part of it-- is delayed.
Apart from cases when the starting point itself is a relevant parameter to be measured (for example in active radar and sonar detection systems), this delay often results in degradation of the other measured signal parameters, especially the amplitudes of the various frequency components. It is therefore convenient to locate the signal starting point so as to be able to optimize the analysis of the data sequence being studied. 

With the exception of wavelets (or multiscale) techniques, which are known to be efficient at locating edges \cite{wav1,wav3} and some parametric models which take into account possible delays \cite{boy}, most methods in the literature compare some measured quantity for a series of contiguous traveling windows, be it power \cite{timfre1,timfre2}, correlation to a library of known waveforms \cite{burst}, or some other quantities \cite{mar}. 

The method we propose instead considers a single window; moreover --when different signal components start at different times-- it locates first the onset time for the signal frequency component with the longest delay, i.e. the first data point where all the signal components are present. This is often the sought for information. By iteration it can then locate the starting points for all other components, should they be needed too.

Our method is based on the analytic properties of the generating function of the data series, which --as we shall see-- depend on the signal delay. More in detail: given an infinite data series $s_{0},s_{1},s_{2},\cdots,s_{n},\cdots$, we can build its 
generating function, or 
``G-transform":
\begin{equation}
G(z)=\sum_{n\geq 0}s_{n}z^{n}.  \label{1}
\end{equation}
In practice, we always have data series that are truncated at order $N-1$, therefore consisting of $N<\infty$ data points. Assuming the total signal to be the discrete sum of a finite number of (damped) harmonic oscillations, we want to estimate the G-transform of the infinite time series by a suitable Pad\'{e} Approximant: a rational function whose Taylor expansion around $z=0$ reproduces the G-transform up to order $N-1$. The problem is that any ratio $[I/L]$ of the order $I$ of the numerator to the order $L$ of the denominator such that $I+L=N-1$ is a candidate Pad\'{e} Approximant; which is the best one?

We know that if all the signals are already present at $n=0$, then --even in the presence of noise-- the best choice is a rational function of type $[M-1/M]$ \cite{noi}, requiring $N=2M$ data points. From its parameters all the (complex) amplitudes and frequencies can be deducted, provided that $M\geq K$, where $K$ is the total number of frequencies in the signal, including complex conjugate pairs.

Things change if some of the frequency components of the signal start at some $n_k$'s larger than zero:   

let's assume we have $K$ signals starting at $n_1,n_2,\cdots,n_K$ with $0\leq n_1\leq n_2\leq \cdots\leq n_K$; as we shall see, the G-transform is then a rational function of the type $[M-1+n_K/M]$.
 
Our first aim is therefore to find the (discrete) time $n_K$ at which the last frequency component starts. To do this, we introduce the U-transform which is the logarithmic derivative of the G-transform:
\begin{equation}
U(z)=\frac{G^{\prime }(z)}{G(z)}. \label{2}
\end{equation}
The poles of the U-transform are the poles and zeros of the G-transform and their residues are $-1$ for the G-transform poles and $+1$ for the zeros; the sum of the residues of the U-transform therefore is $n_K-1$.

If our aim is to find the best Pad\'{e} Approximant for the given sample, we are then done, as $n_K$ is the sought for information. 

If instead we want to find all the starting points $n_k$, $k=1,\cdots K$, we just have to repeat the procedure with a shorter sample of length $2n_K$ to find $n_{(K-1)}$, and so on, until the sample length is reduced to zero.

\section{The theoretical background of the method}\label{theo}

Let us start considering a single discretized signal starting at an unknown time $n_k$:
\begin{eqnarray}
s_{n}=\left\{
\begin{array}{ll}
0 & 0\leq n<n_k \nonumber\\
A e^{i\omega\frac{T}{N}(n-n_k)} \quad \omega=2\pi \nu
+i\alpha,\quad n\geq n_k;
\end{array}
\right.
\end{eqnarray}
the signal can be transient (i.e. damped: $\alpha>0$), of constant amplitude ($\alpha=0$), or even have zero frequency (corresponding to a step function).

The G-transform of this signal reads:
\begin{equation}
G_{k}(z)=z^{n_k}\frac{A_k}{1-ze^{i\omega_k\frac{T}{N}}},\label{5}
\end{equation}
which is a $[n_k/1]$ rational function $z^{n_k}/Q_k^{(1)}(z)$, where $Q_k^{(j)}(z)$ is a polynomial in $z$ of order $j$ and the index indicates that it refers to the signal $k$. Summing $K$ such signals with different starting times, we get

\begin{eqnarray}
G_{signal}(z)=
\Sigma_{k=1}^K G_{k}(z)=
\frac
{\Sigma_{k=1}^K z^{n_k}A_k\Pi_{j=1,K}^{j\neq k}Q_j^{(1)}(z)}
{\Pi_{j=1,K}Q_j^{(1)}(z)}=
\frac
{\Sigma_{k=1}^K z^{n_k}Q_{k}^{(K-1)}(z)}
{Q_{signal}^{(K)}(z)}=
\frac
{P_{signal}^{(K-1+n_K)}(z)}
{Q_{signal}^{(K)}(z)},
\end{eqnarray}
where $n_K = \max_{k=1,K} (n_k)$ and we have indicated with $P$ the polynomial at the numerator and with $Q$ the one at the denominator.

Adding noise does not change the the substance of the above picture, as it means to add an equal number $H$ of poles and zeros (Froissart doublets \cite{dou1,dou2,dou3,dou4}) so that the complete G-transform reads:
\begin{eqnarray}
G_{total}(z)=
\frac
{P_{signal}^{(K-1+n_K)}(z)}
{Q_{signal}^{(K)}(z)}+
\frac
{P_{noise}^{(H)}(z)}
{Q_{noise}^{(H)}(z)}=
\frac
{P_{total}^{(K+H-1+n_K)}(z)}
{Q_{total}^{(K+H)}(z)}=
\frac
{\Pi_{i=1}^{K+H-1+n_K}(z-z_i)}
{\Pi_{j=1}^{K+H}(z-p_j)},
\end{eqnarray}
where we have respectively called $z_i$ and $p_j$ the zeros and poles of $G_{total}(z)$, which is a $[(K+H-1+n_K)/(K+H)]$ rational function, thus requiring $N=2K+2H+n_k$ data points to be reconstructed by Pad\'{e} Approximant.

Now that we know the form of the G-transform, we can calculate the form of the corresponding U-transform, eq. \ref{2}:
\begin{eqnarray}
U(z)=\Sigma_{j=1}^{K+H}\frac{-1}{z-p_j}+\Sigma_{1=1}^{K+H-1+n_K}\frac{+1}{z-z_i}. \label{3}
\end{eqnarray}
From this last formula, it is evident that $U(z)$ is a rational function of type $[M-1/M]$ with $M=2K+2H+n_K-1$; the $M$ residues of its poles can only assume the values $\pm 1$ and --most important-- $n_K-1$ is equal to the number of poles with residue $+1$ minus the number of poles of residues $-1$  i.e. the sum of the residues itself, as we stated in the Introduction.

We conclude this section by noting that the above argument works not only for signals starting at $n_1,n_2,\cdots,n_K$, but also for signals peaking at those points.
This is because the growing part of each signal ends at the starting point of the decaying part, and therefore its Z-transform is a polynomial. More in detail: let us consider a signal of the form
\begin{eqnarray}
s_{n}=A e^{i\omega\frac{T}{N}|n-n_k|}\quad \omega=2\pi \nu
+i\alpha,\quad \alpha>0; 
\end{eqnarray}
it's G-transform will be
\begin{equation}
G_{k}^{sym}(z)=A_k e^{i\omega_k\frac{T}{N}n_k}\Sigma_{n=0}^{n_k-1} \left({z e^{-i\omega_k\frac{T}{N}}}\right)^n
+A_k z^{n_k}\Sigma_{n=0}^{\infty} \left({z e^{i\omega_k\frac{T}{N}}}\right)^n.
\end{equation}
The first addendum is a polynomial in $z$ of order $n_k-1$, while the second addendum is again given by eq. \ref{5}; we can therefore write
\begin{equation}
G_{k}^{sym}(z)=P^{(n_k-1)}(z)
+z^{n_k}\frac{A_k}{1-ze^{i\omega_k\frac{T}{N}}}=
\frac{Q_k^{(1)}(z)P^{(n_k-1)}(z)+A_k z^{n_k}}{Q_k^{(1)}(z)},
\end{equation}
which again is a $[n_k/1]$ rational function.

\section{Some practical details for the numerics}\label{6}

The first step of our method is to compute the Taylor series of $U(z)$ from the one (eq. \ref{1}) defining $G(z)$, so as to be able to then build its Pad\'{e} Approximant, from which to get the residues of its poles, and finally $n_K$.

From Ref. \cite{ref46} (section $0.31$), the coefficients $c_n/a_0$ of the Taylor expansion of the ratio of two expansions
\begin{eqnarray}
U(z)=\frac{\Sigma_{n=0}^{\infty}b_n z^n}{\Sigma_{n=0}^{\infty}a_n z^n}
\end{eqnarray}
are given by the recurrence formula
\begin{eqnarray}
c_n=b_n-\frac{1}{a_0}\Sigma_{k=1}^{n} c_{n-k} a_k
\end{eqnarray}
in our case the coefficients of the series at the denominator are $a_n=s_n$ from eq. \ref{1}; those of the series at the numerator are $b_n=(n+1)s_{n+1}$. It is therefore convenient to define $\alpha_n=s_n/s_0$ and $\gamma_n=c_n/s_0$ so as to have a recurrence formula for the coefficients themselves:
\begin{eqnarray}
\gamma_n=(n+1)\alpha_{n+1}-\Sigma_{k=1}^{n} \gamma_{n-k} \alpha_k. \label{4}
\end{eqnarray}

From eq. \ref{3}, we have seen that the Pad\'{e} for $U(z)$ has to be be of type $[M-1/M]$, with $M=(2K+2H+n_K-1)$ poles. We therefore need a $2M$ term expansion for $U(z)$. We'll have to get to $\gamma_{2M-1}$, or $\alpha_{2M}$ in the recurrence formula eq. \ref{4}, for which we need $N=2M+1$ data points from the original series, as opposed to the $M+1$ points only, needed to reconstruct $G(z)$.

The maximum possible value for $n_K$ corresponds to the case when the data series contains only the signal starting at $n_K$, meaning $K=1$ and $H=0$; therefore, $M=n_K^{max}+1$ and $N=2M+1=2n_K^{max}+3$, which gives us the bound
\begin{eqnarray}
n_K<\frac{N-3}{2}
\end{eqnarray}
for the possible delays detectable from a truncated sequence of  $N$ data points.

%The observed erratic behavior of the U-transform when the k-shift reaches one half of the data number is explained by the following remark:
%Let $S$ be the number of signals starting at an unknown time $k_1$.
%Let $V$ be the number of stationary signals 
%The Pad\'{e} to use is the $[L-1/L]$ one. U-transform, NOT Z-transform. Note that Daniel Vrinceanu used an $[L/L]$ one. I do not think it has any practical importance.
%The Pad\'{e} of the U-transform used is a Pad\'{e} $[L-1/L]$ one with $L=4V+4S+k_1 -1$. The number of data in the U-Taylor series required for its stable reconstruction is $2L$, that is:
%$8V+8S+2 k_1 -2$ or for the Z-Taylor series:
%$N_{min}= 8V+8S+2 k_1 -1$
%So
%$N>8V+8S+2 k_1$
%This number is  $k_1$ dependent and increases linearly with $k_1$. The bigger $k_1$, the bigger the number of data implied. The inequality of consistence is certainly violated when $k_1>N/2$.

\section{Numerical tests}

We present here some numerical examples as a test of the method described above. 

\begin{figure}[htbp]
\centering\epsfig{file=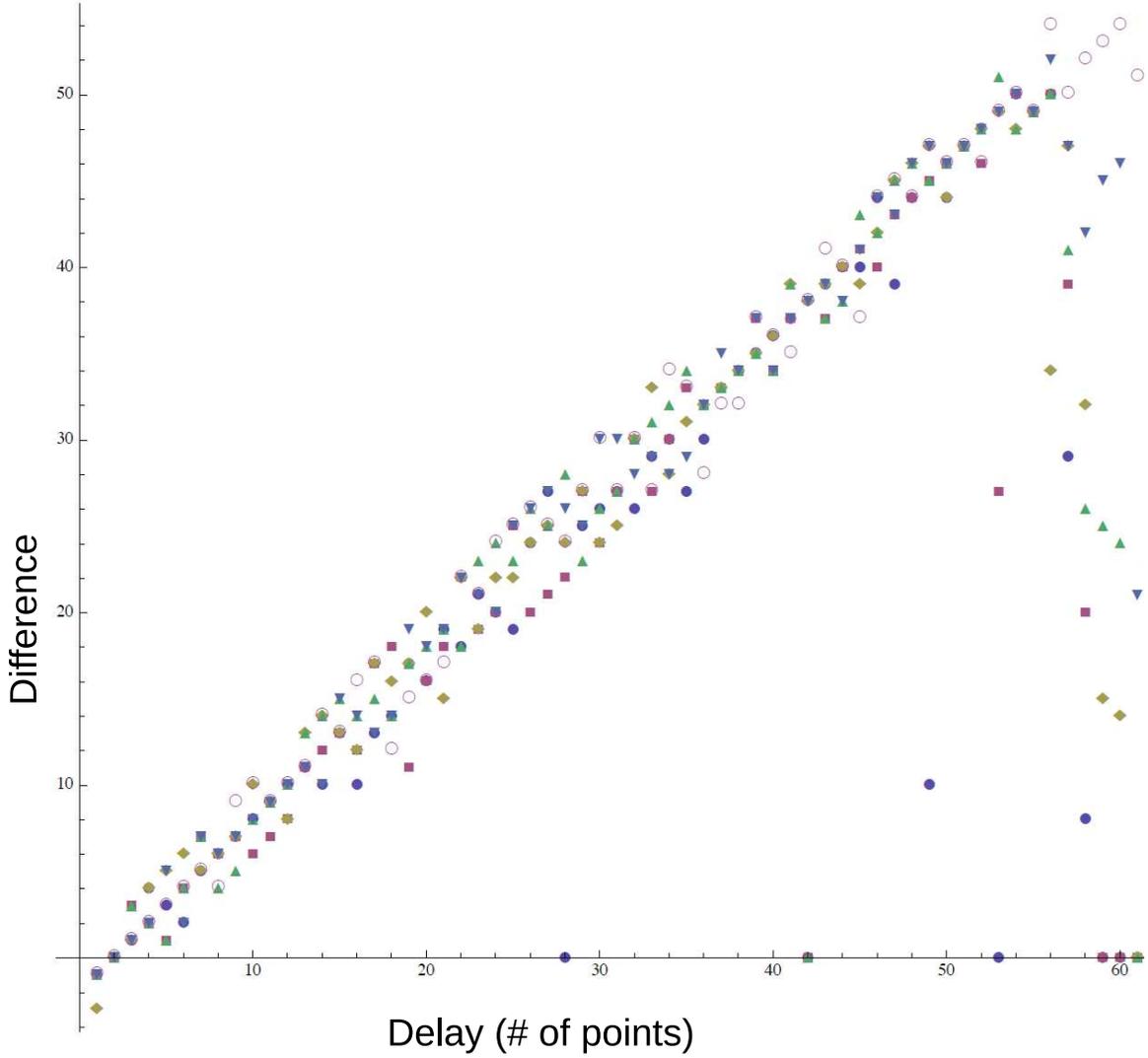,width=0.9\linewidth}
\caption{The difference between the number of residues around $+1$ (within a radius $r= 100\cdot \mu$) and
the number of residues around $-1$ (again within a radius $r= 100\cdot \mu$), as a function of the delay,
for various levels of noise: $\mu=10^{-8}$ (full circles), $\mu=10^{-7}$ (full squares), $\mu=10^{-6}$ (diamonds), $\mu=10^{-5}$ (up triangles), $\mu=10^{-4}$ (down triangles), $\mu=10^{-3}$ (empty circles). The parameters are: $129$ data points $f_1=120/1024$, $f_2=280/1024$, $\gamma=280/1024$. The calculation was performed using Mathematica\copyright with $140$ digits of precision.}
\label{fig1}
\end{figure}

Our first example considers a signal of the form
\begin{eqnarray}
s_{n}=\left\{
\begin{array}{ll}
\cos(2*\pi*f_1*n) & 0\leq n<n_1 \nonumber\\
\cos(2*\pi*f_1*n)+ e^{-i\gamma(n-n_1)}\cos(2*\pi*f_2*(n-n_1)) & n\geq n_1
\end{array}
\right.
\end{eqnarray}
to which random noise uniform in $[-\mu,\mu]$ is added. 
In Figure \ref{fig1} we show the difference between the number of residues around $+1$ (within a radius $r= 100\cdot \mu$) and the number of residues around $-1$ (again within a radius $r= 100\cdot \mu$), as a function of the delay $n_1$, for various levels of noise between $10^{-8}$ and $10^{-3}$. The plot is in agreement with our statement  in section \ref{theo} that the delay equals said difference. As expected, the linear dependence of the difference on the delay fails at about $N/2$.

\begin{figure}[htbp]
\centering\epsfig{file=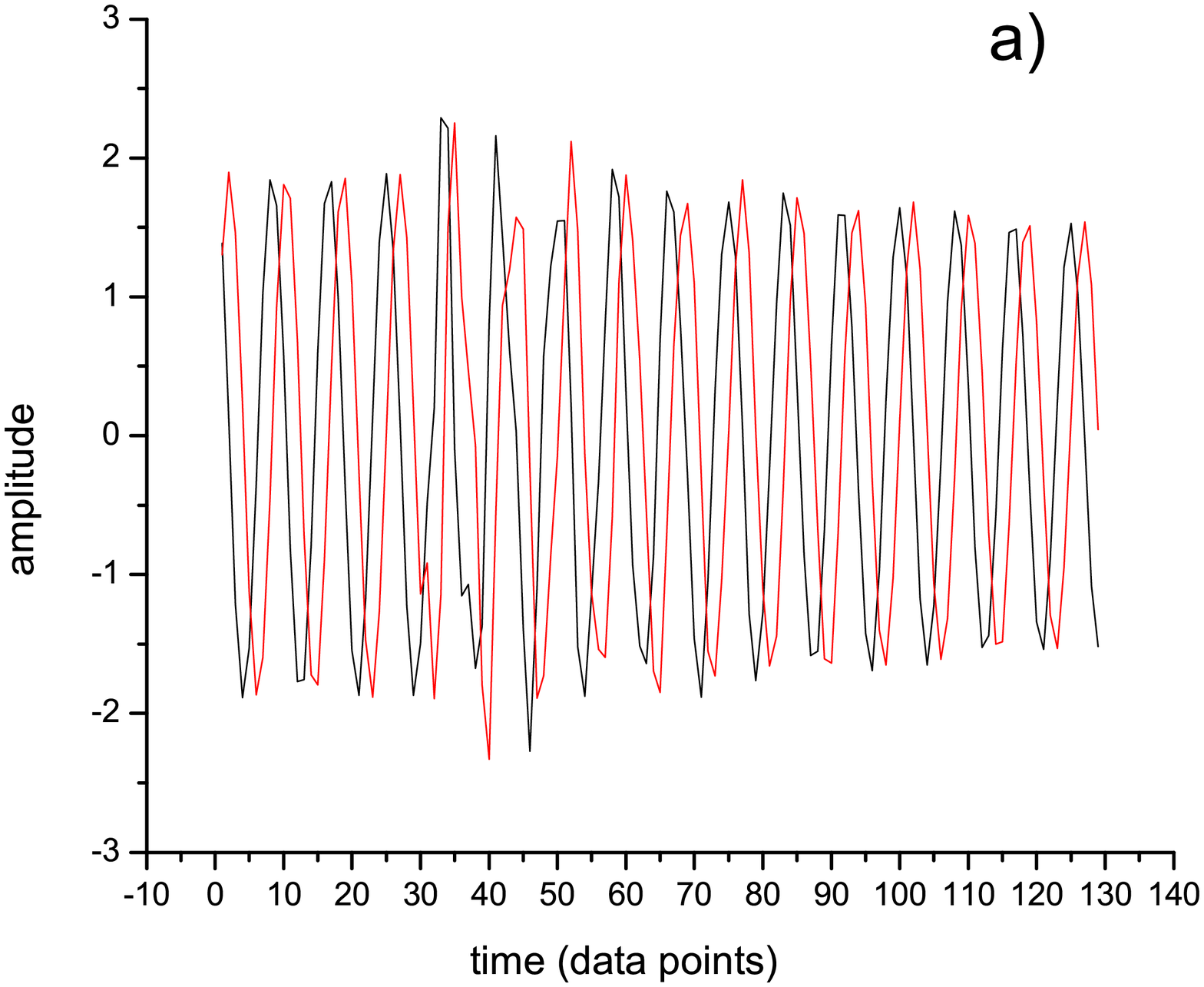,width=0.5\linewidth}
\centering\epsfig{file=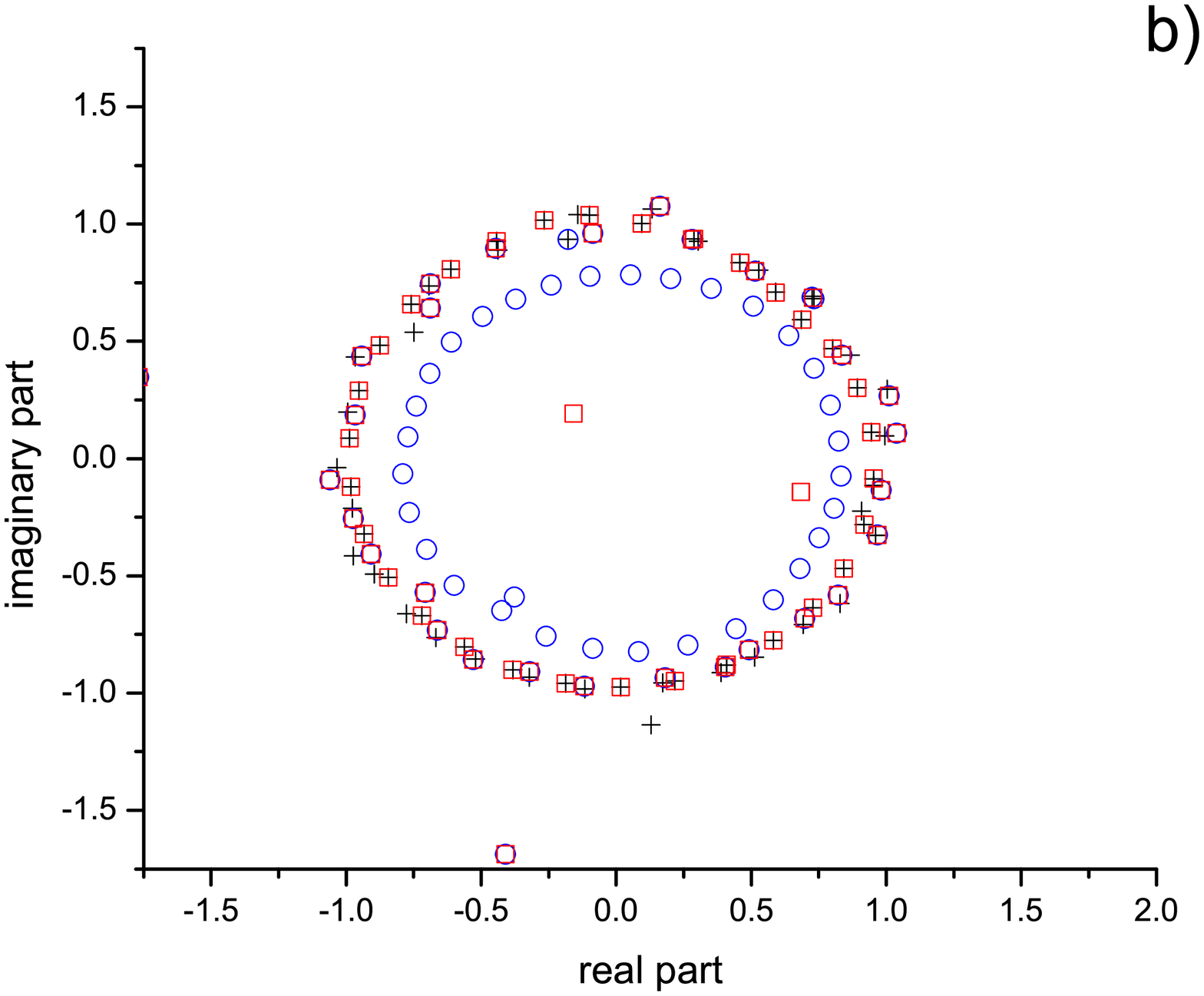,width=0.5\linewidth}
\centering\epsfig{file=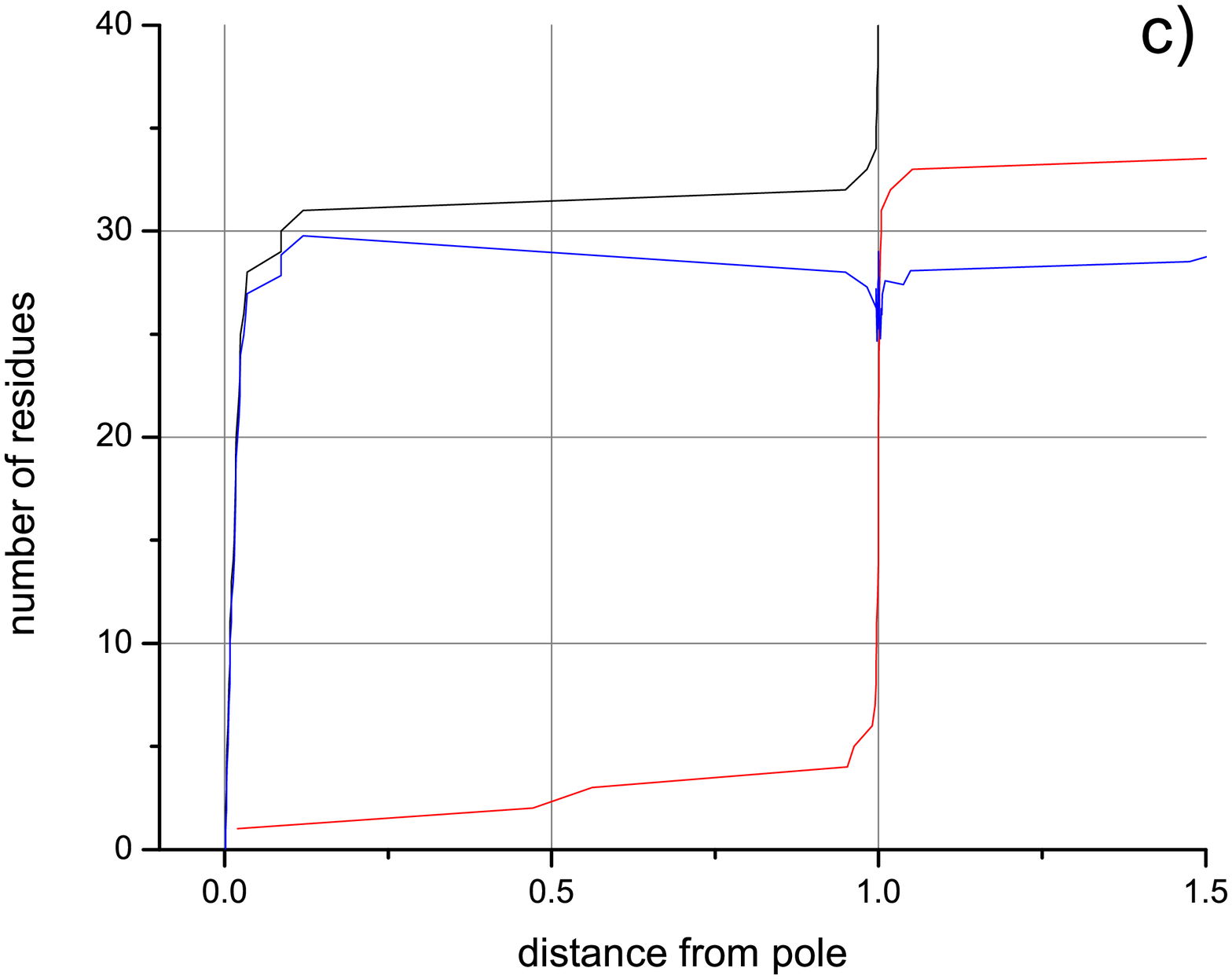,width=0.5\linewidth}
\caption{ a) The real (black line) and imaginary (red line) parts of the signal given by eq.\ref{esponenti} with $\omega_1=0.75$,  $\omega_2=0.76$, $\omega_3=1.76$, and $\gamma=0.05$. $129$ data points. b) The poles of the Pad\'{e} approximant to $U(z)$ (crosses) and the poles and zeros of of the $[n-1/n]$ Pad\'{e} approximant to $G(z)$ (circles and squares, respectively). c) The cumulative number of residues as a function of the distance from $+1$ (black line) and from $-1$ (red line), together with their difference (blue line). The calculation was performed using Octave\copyright.}
\label{fig2}
\end{figure}

Our next example considers the case of three signals, one of which one is delayed: 
\begin{eqnarray}
s_{n}=\left\{
\begin{array}{ll}
e^{i\omega_1 n} + 0.9e^{i\omega_2 n} & 0\leq n<n_1 \nonumber\\
e^{i\omega_1 n} + 0.9e^{i\omega_2 n}+ e^{-\gamma(n-n_1)}e^{\omega_3*(n-n_1)} & n\geq n_1. \label{esponenti}
\end{array}
\right.
\end{eqnarray}
In this case, the noise we add is a complex Gaussian random noise with variance $\sigma=10^{-4}$ in both real and imaginary part. 
In Figure \ref{fig2}.a we show the signal for $n_1=30$: as the amplitudes of the three signals are comparable, the beginning of the delayed signal is barely visible. Figure \ref{fig2}.b shows the poles of the Pad\'{e} approximant to $U(z)$ (crosses) and the poles and zeros of of the $[n-1/n]$ Pad\'{e} approximant to $G(z)$ (circles and squares, respectively): clearly something is wrong with the assumption that $G(z)$ is a rational function of the $[n-1/n]$ type: instead of having the $G(z)$ poles distributed around the unit circle, $31$ poles form a circle inside the unit circle (two more are close but not on the circle itself). Finally Figure \ref{fig2}.c shows the cumulative number of residues as a function of the distance from $+1$ (black line) and from $-1$ (red line), together with their difference (blue line): the latter shows a maximum of $30$ at a distance of about $0.1$. 

\begin{figure}[htbp]
\centering\epsfig{file=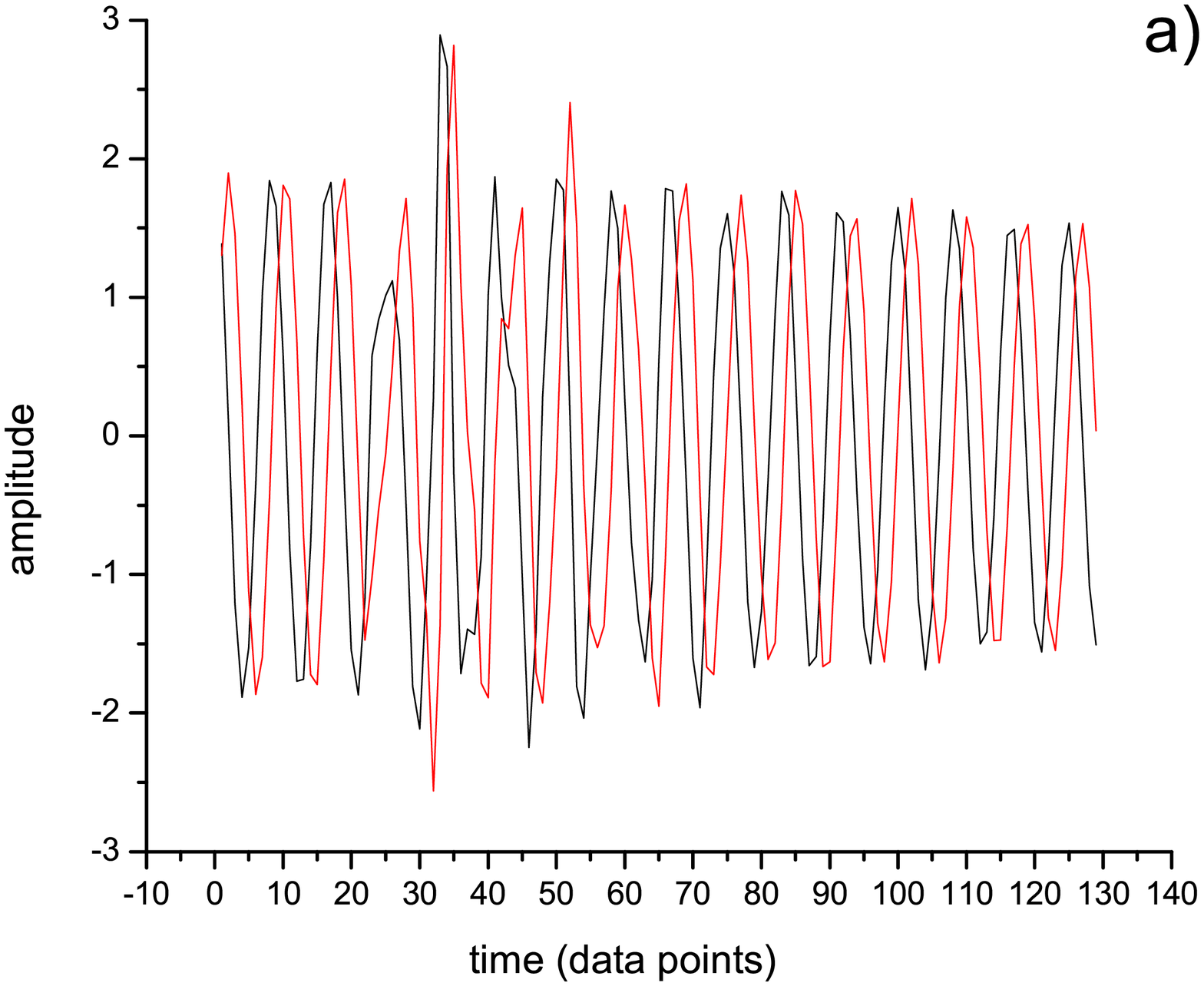,width=0.5\linewidth}
\centering\epsfig{file=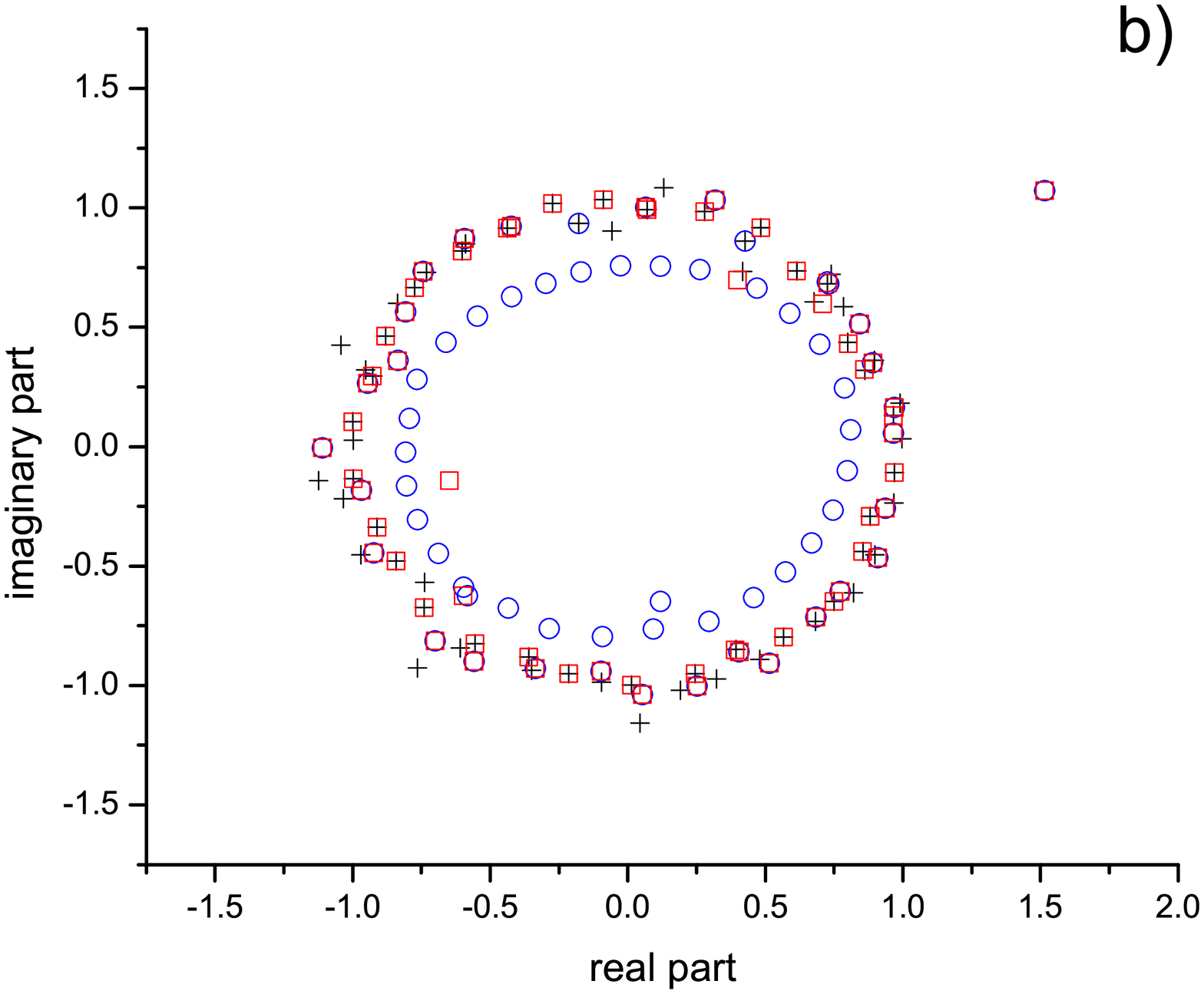,width=0.5\linewidth}
\centering\epsfig{file=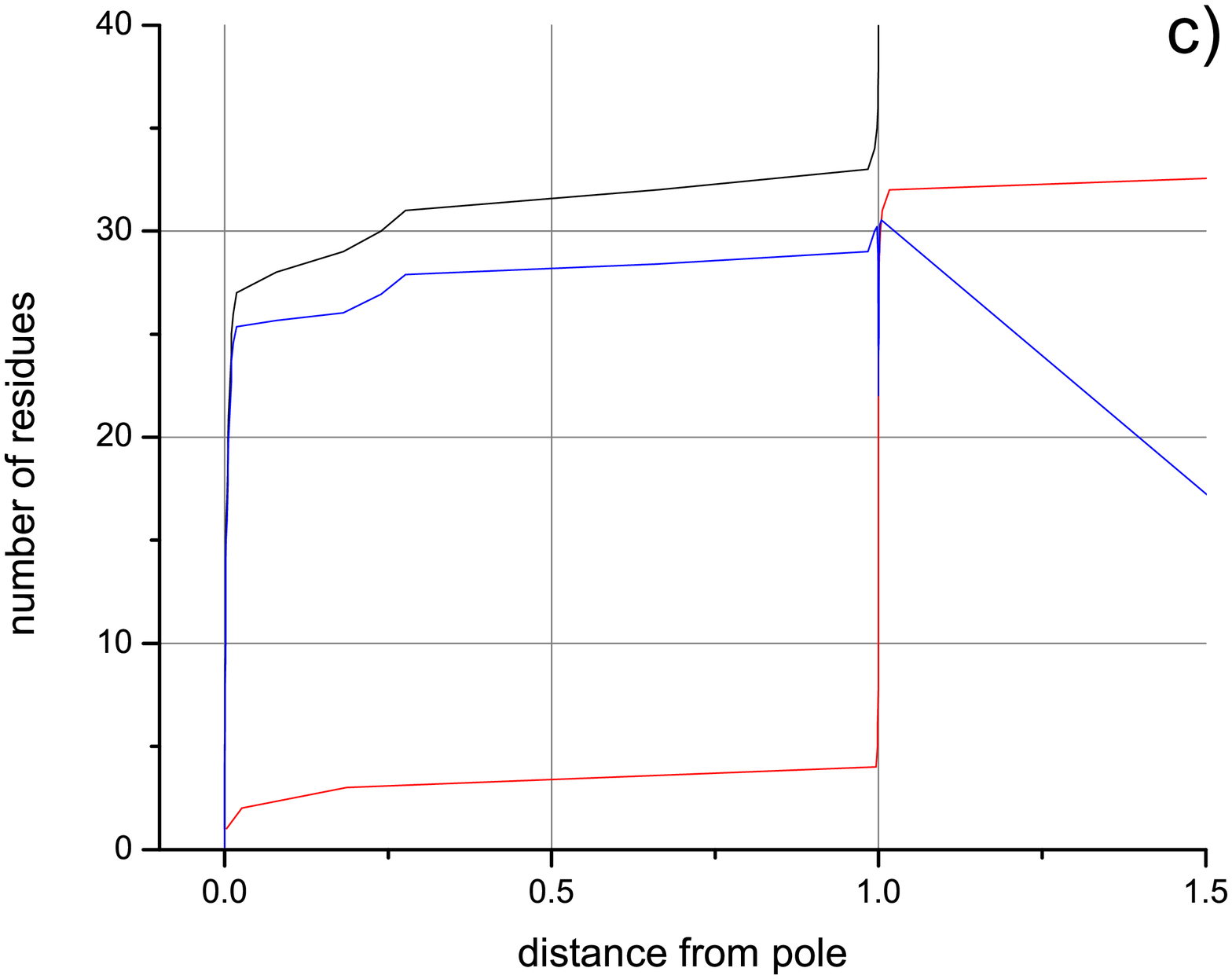,width=0.5\linewidth}
\caption{ a) The real (black line) and imaginary (red line) parts of the signal given by eq.\ref{due} with $\omega_1=0.75$, $\omega_2=0.76$, $\omega_3=1.11$, $\gamma_3=0.04$, $\omega_4=1.76$, and $\gamma_4=0.05$. $129$ data points. b) The poles of the Pad\'{e} approximant to $U(z)$ (crosses) and the poles and zeros of of the $[n-1/n]$ Pad\'{e} approximant to $G(z)$ (circles and squares, respectively). c) The cumulative number of residues as a function of the distance from $+1$ (black line) and from $-1$ (red line), together with their difference (blue line). The calculation was performed using Octave\copyright.}
\label{fig3}
\end{figure}

To check our claim that the difference between the number of poles with residue $+1$ and those with residues $-1$ equals the longest delay, we now pass to the case of two delayed signals:
\begin{eqnarray}
s_{n}=\left\{
\begin{array}{ll}
e^{i\omega_1 n} + 0.9e^{i\omega_2 n} & 0\leq n<n_1 \nonumber\\
e^{i\omega_1 n} + 0.9e^{i\omega_2 n}+ e^{-\gamma(n-n_1)}e^{\omega_3*(n-n_1)} & n_1\leq n < n_2 \nonumber\\
e^{i\omega_1 n} + 0.9e^{i\omega_2 n}+ e^{-\gamma_3(n-n_1)}e^{\omega_3*(n-n_1)}+ e^{-\gamma_4(n-n_2)}e^{\omega_4*(n-n_2)} & n\geq n_2 \label{due}
\end{array}
\right.
\end{eqnarray}
to which complex Gaussian random noise with variance $\sigma=10^{-4}$ in both real and imaginary part is again added. 
In Figure \ref{fig3}.a we show the signal for $n_1=22$ and $n_2=30$. Figure \ref{fig3}.b shows the poles of the Pad\'{e} approximant to $U(z)$ (crosses) and the poles and zeros of of the $[n-1/n]$ Pad\'{e} approximant to $G(z)$ (circles and squares, respectively): again the assumption that $G(z)$ is a rational function of the $[n-1/n]$ type produces a circle of poles inside the unit circle. Figure \ref{fig3}.c shows the cumulative number of residues as a function of the distance from $+1$ (black line) and from $-1$ (red line), together with their difference (blue line). This time the difference only reaches $26$ at a distance of about $0.1$ and only gets to $30$ for a distance close to $1$. 

We finally consider the case when the delayed signal is present all the time, but peaks at $n=n_1-1$:
\begin{eqnarray}
s_{n}=\left\{
\begin{array}{ll}
e^{i\omega_1 n} + 0.9e^{i\omega_2 n}+ e^{-\gamma(n_1-n)}e^{\omega_3*(n_1-n)} & 0\leq n\leq n_1 \nonumber\\
e^{i\omega_1 n} + 0.9e^{i\omega_2 n}+ e^{-\gamma(n-n_1)}e^{\omega_3*(n-n_1)} & n > n_1. \label{picco}
\end{array}
\right.
\end{eqnarray}
Again complex Gaussian random noise with variance $\sigma=10^{-4}$ in both real and imaginary part is added and we consider $n_1=30$. The results are very similar to those of Figure \ref{fig2}; as expected the maximum difference is $29$.

\begin{figure}[htbp]
\centering\epsfig{file=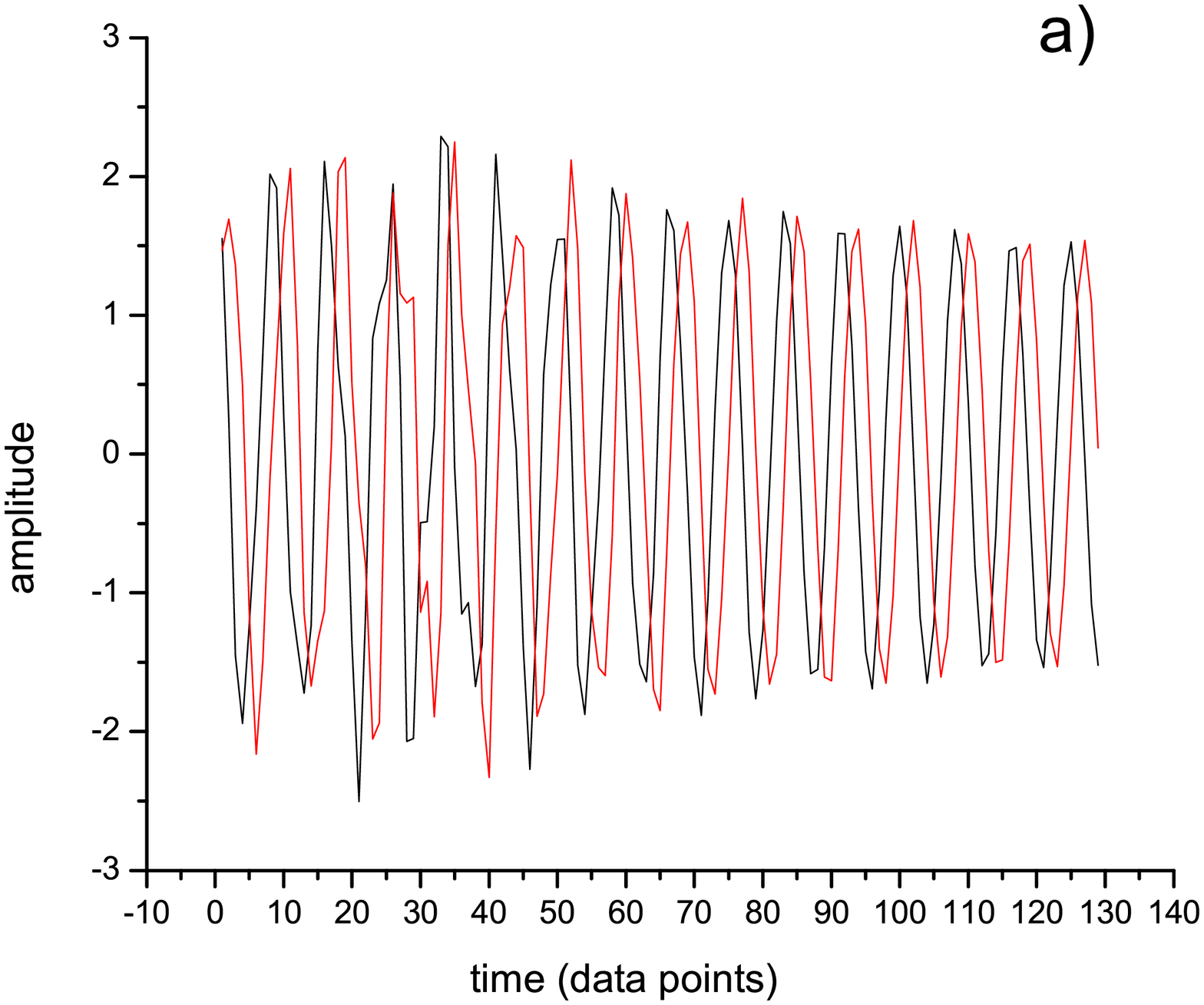,width=0.5\linewidth}
\centering\epsfig{file=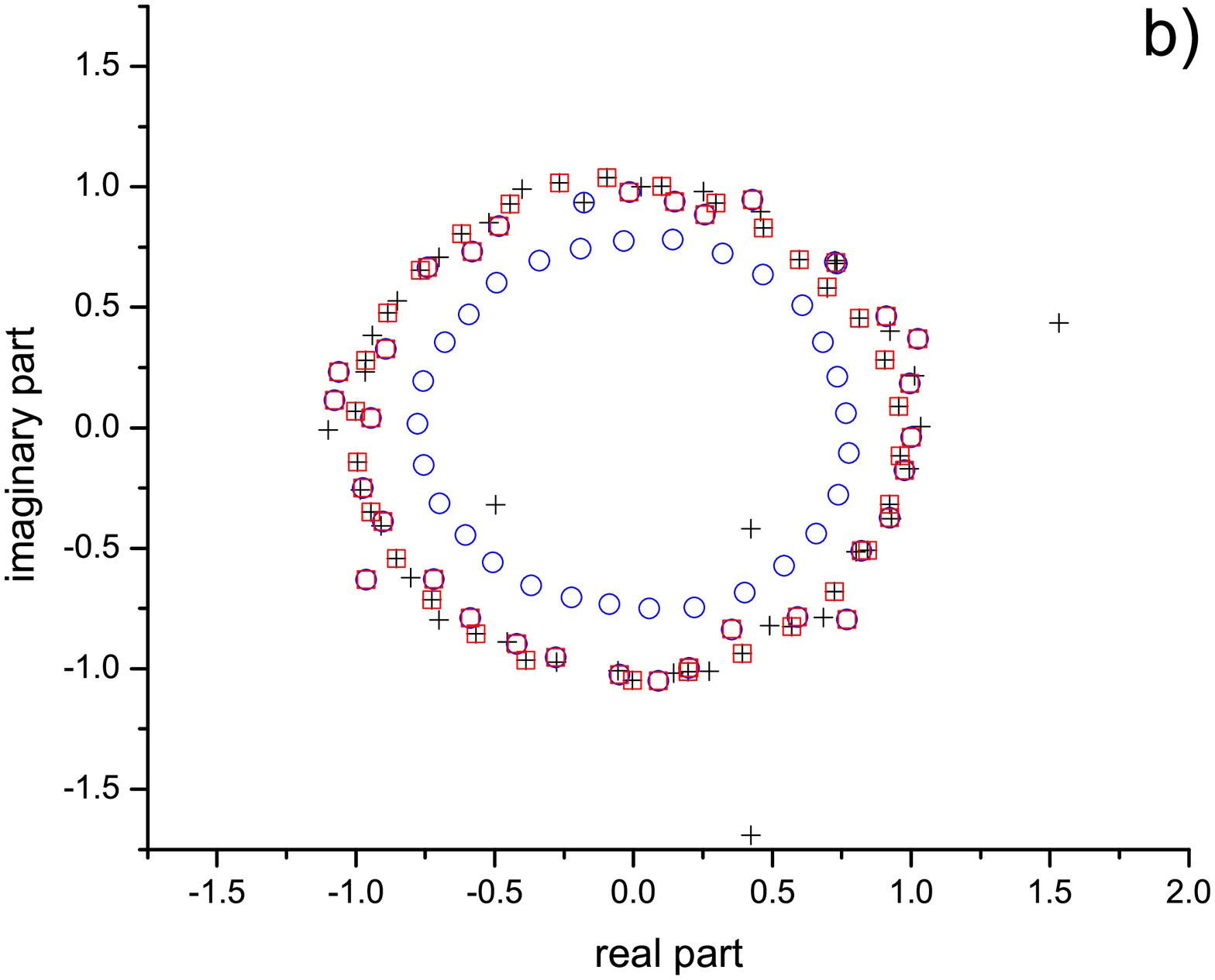,width=0.5\linewidth}
\centering\epsfig{file=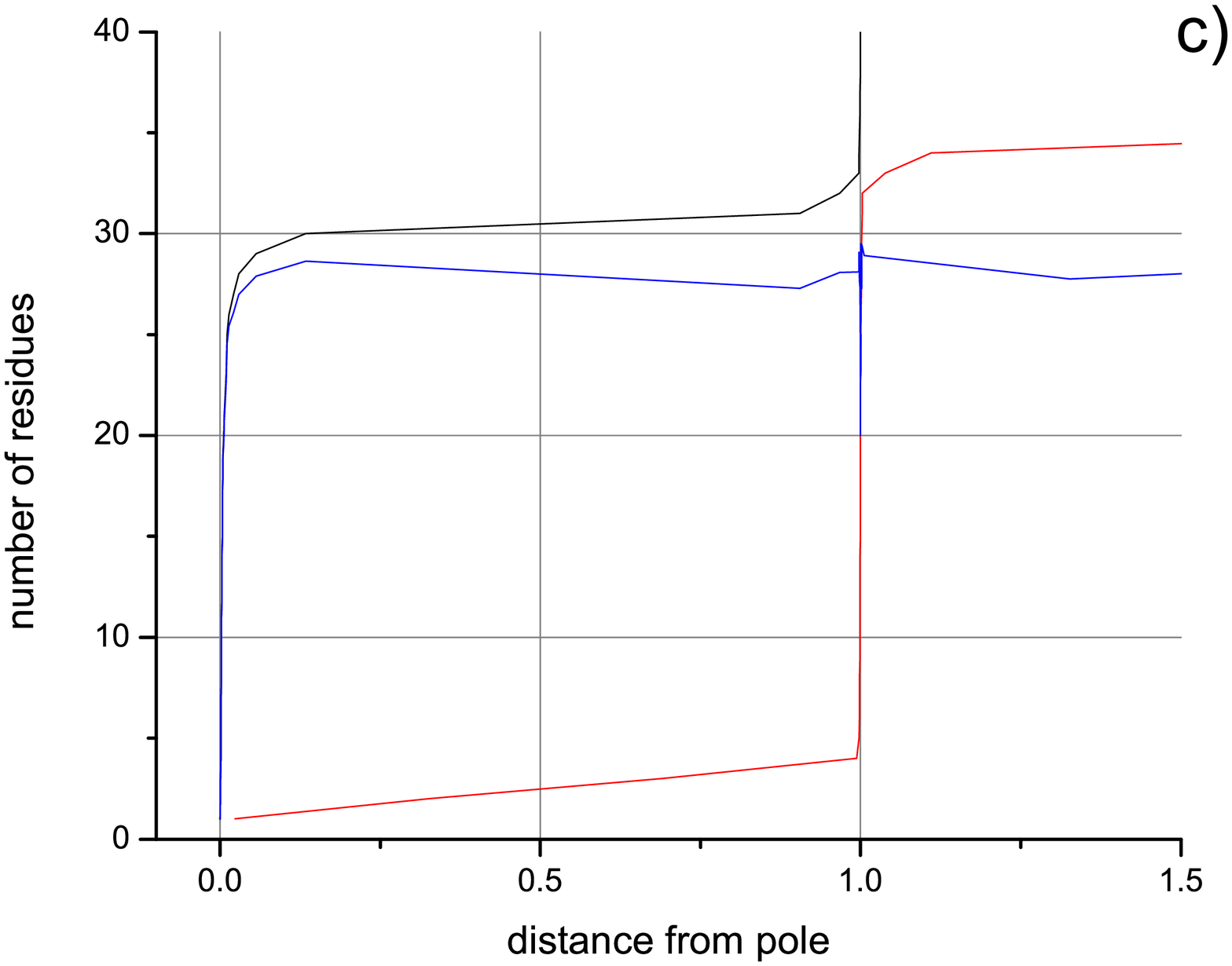,width=0.5\linewidth}
\caption{ a) The real (black line) and imaginary (red line) parts of the signal given by eq.\ref{picco} with $\omega_1=0.75$,  $\omega_2=0.76$, $\omega_3=1.76$, and $\gamma=0.05$. $129$ data points. b) The poles of the Pad\'{e} approximant to $U(z)$ (crosses) and the poles and zeros of of the $[n-1/n]$ Pad\'{e} approximant to $G(z)$ (circles and squares, respectively). c) The cumulative number of residues as a function of the distance from $+1$ (black line) and from $-1$ (red line), together with their difference (blue line). The calculation was performed using Octave\copyright.}
\label{fig4}
\end{figure}

\section{Conclusions}

We have shown that the information about the starting points in time of a number of delayed signals can be directly retrieved from the residues of the poles of the U-transform of the signal: they can only be $\pm 1$ and their sum plus one equals the longest delay among those of the signal's frequency components. We have then described how an iterative algorithm using this property can be implemented in practice to locate the starting points in time of all the frequency components of a signal. Numerical examples confirm the feasibility and stability of the proposed scheme.


\begin{thebibliography}{6}

%\bibitem{ecg} N. Vasudha; ``Detection of Discontinuity in ECG Using Wavelet Transform", Applied Mathematical Sciences, Vol. 6, 2012, no. 117, 5835 – 5840.

\bibitem{wav1} M. Sharifzadeh, F. Azmoodeh, and C. Shahabi; ``Change Detection in Time Series Data Using Wavelet Footprints", C. Bauzer Medeiros et al. (Eds.): SSTD 2005, LNCS 3633, pp. 127–144, 2005.

%\bibitem{wav2} A. Antoniadis and I. Gijbels; ``Detecting abrupt changes using wavelet methods", ........

\bibitem{wav3} S. Mallat and W. L. Hwang; ``Singularity detection and processing with wavelets",
IEEE Trans. Inf. Th, 38:617–643, 1992.

\bibitem{boy} R. Boyer and K. Abed-Meraim, “Damped and delayed model for
transient modeling,” IEEE Trans. Signal Process., vol. 53, no. 5, pp.
1720–1730, May 2005.

%BURST TIME FREQUENCY METHOD :

\bibitem{timfre1} W. Anderson and R. Balasubramanian, Phys. Rev. D {\bf 60}, 102001 (1999).

\bibitem{timfre2} W.G. Anderson, P.R. Brady, J.D.E. Creighton, E.E. Flanagan, Phys. Rev. D {\bf 63}, 042003 (2001).

%MATCH FILTERING FOR CHIRPS : 
\bibitem{burst} B. A. Allen, W. G. Anderson, P. R. Brady, D. A. Brown, and J. D. E. Creighton, arXiv:gr-qc/0509116 (2005).
%B. Allen, W. G. Anderson, P. R. Brady, D. A. Brown and J. D. E. Creighton, arXiv:gr-qc/0509116.

%\bibitem{boy} R. Boyer and K. Abed-Meraim; ``Asymptotic Performance for Delayed Exponential Process", IEEE Transactions on Signal Processing, vol. 55, no. 6, June 2007.

\bibitem{mar} S. K. Marks and R. Gonzalez; ``Algorithms for improved sinusoidal track onset localisation", in ``Proceeding of Signal and Image Processing conference" M.H. Hamza ed. (ACTA press, Calgary, 2004).

\bibitem{noi} D. Bessis and L. Perotti; {\it Universal analytic properties of noise: introducing the J-matrix formalism}, J. Phys. A {\bf 42} (2009) 365202.

%\bibitem{stei}H. Steinhaus;  {\it Uber die Wahrscheinlichkeit dafuer dass
%der Konvergenzkreis einer Potenzreihe ihre natuerliche Grenze ist}, Math. Z. {\bf 31} (1929) pp. 408-416.

%\bibitem{fro} M. Froissart, Approximation de Pad\'{e}: Application \`{a} la Physique des Particules \'{e}l\'{e}mentaires
%in: J. Carmona, M. Froissart, D.W. Robinson, D. Ruelle (Eds.), Recherche Coop\'{e}rative sur Programme (RCP), Centre National de la Recherche Scientifique (CNRS),Strasbourg, L 8470, No. 25, 1969, p. 1.

%\bibitem{carasau} D. Bessis, L. Perotti, and D. Vrinceanu; {\it Noise in the complex plane: open problems},
%Numer. Algor. {\bf 62} (2013) pp. 559-569.

\bibitem{dou1}J. Gilewicz and Truong-Van; {\it Froissart --Doublets in
Pad\'{e} Approximants and Noise} in {\it Constructive Theory of
Functions 1987}, Bulgarian Academy of Sciences, Sofia, 1988. pp.
145-151

\bibitem{dou2}J-D. Fournier, G. Mantica, A. Mezincescu, and D. Bessis; {\it %
Universal statistical behavior of the complex zeros of Wiener transfer
functions}, Europhysics Letter 22, 325-331 (1993)

\bibitem{dou3}J-D. Fournier, G. Mantica, A. Mezincescu, and D. Bessis; {\it %
Statistical Properties of the zeros of the transfer functions in signal
processing}, in {\it Chaos and Diffusion in Hamiltonian Systems}, D. Benest
and C. Froeschle eds. (Editions Fronti\`{e}res, 1995).

\bibitem{dou4}D. Bessis; {\it Pad\'{e} Approximations in noise filtering},
International Congress on Computational and Applied Mathematics. Journal of
Computational and Applied Mathematics 66, 85-88 (1996)

\bibitem{ref46} Gradshtenyn I S and Ryzhik I M 1980 {\it Table of Integrals, Series, and Products}, Academic Press.

\end{thebibliography}
\end{document}